# Draft of paper submitted to Icarus
# to be published with open access

**Seismicity on Tidally Active Solid-Surface Worlds**

Hurford, T.A.[1], Henning, W.G.[1,2], Maguire, R.[3], Lekic, V.[3], Schmerr, N.[3], Panning, M.[4], Bray, V.J.[5], Manga, M.[6], Kattenhorn, S.A.[7], Quick, L. C.[8], Rhoden, A.R.[9,10]

[1]Planetary Geology, Geophysics and Geochemistry Laboratory, NASA Goddard Space Flight Center, Greenbelt, Maryland, USA. [2]University of Maryland, College Park, Maryland, USA. [3]Department of Geology, University of Maryland, College Park, Maryland, USA. [4] Jet Propulsion Laboratory, California Institute of Technology, Pasadena, California, USA. [5]Lunar and Planetary Laboratory, University of Arizona, Tucson, Arizona, USA. [6]University of California, Berkeley, California, USA. [7]University of Alaska Anchorage, Anchorage, Alaska, USA. [8]Center for Earth and Planetary Studies, National Air and Space Museum, Smithsonian Institution, Washington, DC, USA.[9]Arizona State University, Tempe, Arizona, USA. [10]SouthWest Research Institution, Boulder, Colorado, USA.

## Abstract

Tidal interactions between planets or stars and the bodies that orbit them dissipate energy in their interiors. The energy dissipated drives internal heating and a fraction of that energy will be released as seismic energy. Here we formalize a model to describe the tidally-driven seismic activity on planetary bodies based on tidal dissipation. To constrain the parameters of our model we use the seismic activity of the Moon, driven by tidal dissipation from the Earth-Moon interactions. We then apply this model to survey the amount of seismic energy release and largest seismic events on other moons in our Solar System and exoplanetary bodies. We find that many moons in the Solar System should be more seismically active than the Earth's Moon and many exoplanets should exhibit more seismic activity than the Earth. Finally, we examine how temporal-spatial variations in tidal dissipation manifest as variations in the locations and timing of seismic events on these bodies.

**1 Introduction**

The seismic activity of tidally-driven planets and moons outside the Earth-Moon system is unknown, and even the tidally-driven activity of Earth's Moon remains poorly constrained. Notwithstanding, we do know that the Moon exhibits seismic activity that is linked to tidal interactions with the Earth (e.g., Latham et al., 1971; Lammlein, 1977; Toksoz et al., 1977; Nakamura, 1977), we have observed tidal control of activity from fractures on Enceladus (Hurford et al., 2007; Hedman et al. 2013; Nimmo et al., 2014) and we can observe complex tectonic fabrics on many tidally-influenced bodies in our Solar System. In fact, the Moon's tides also influence seismicity on all types of plate boundaries on Earth, e.g., including divergent plate boundaries (e.g., Tolstoy et al, 2002), non-volcanic tremor at convergent boundaries (e.g., Rubinstein et al., 2008), and along transform boundaries (e.g., van der Elst et al., 2016).

Previously, seismic activity on planetary objects has been studied through descriptions of near-surface stresses, which are often tidally induced. This approach has a few inherent weaknesses, which include:

- Any stress method focusing on one stress type (e.g. tensile or shear failure) may fail to account for events triggered by other stress types or combinations of stresses not considered.
- Stress methods depend greatly on the orientation of faults. For many tidally active targets of interest, surface fault distributions remain unknown or unmapped at the scales needed for seismic predictions. On Europa, only the largest faults are mapped at global scales. These global faults do not capture the smaller scale fault systems, do not represent fault patterns just below the surface or at mid-shell depths, nor do they account for the dip angle of faults as they descend through the ice shell.

Here, we explore an alternative that seeks to overcome the weaknesses inherent in previous methods that constrain seismicity based on the consideration of stress alone: instead of trying to relate seismicity to stress patterns and fault distributions, we attempt to link the total tidally dissipated energy to expected seismic activity. "Energy methods" in many fields (certainly for deformation, bending, or crushing in engineering) are often simpler than trying to resolve all microscale stress phenomena, particularly for complex phenomena involving irregular geometry.

As in most such energy methods, reliability of the method is not contingent upon knowing the exact path of all energy cascades throughout the system, but instead upon having broad-level and generalizable observational constraints. Scaling overall seismic activity level to tidal dissipation energy has been proposed for Europa (Panning et al., 2018), but this approach did not re-analyze the lunar seismic record to constrain tidally-driven seismic activity, was not generalized to other tidally active worlds, nor did it account for spatiotemporal variation of tidal dissipation energy deposition.

In this study, we outline a method to estimate the size and frequency distribution of seismic events on tidally active worlds. We show how tides may affect the timing and location of events occurring on these bodies. In developing this framework, we use the Moon to constrain links between tidal dissipation and tidally driven seismic activity. Finally, we detail interesting test cases for Io, Europa, and examples of a tidally active terrestrial-class exoplanets, such as TRAPPIST-1b.

## 2 Tidal dissipation and seismic energy

Here we develop a method to estimate tidally-driven seismicity by linking it to tidal dissipation within tidally active worlds. While much about the seismicity of these worlds remains unknown, constraints on tidal dissipation provide a starting point for assessing possible seismicity rates.

Tidally-active bodies are those which experience a large amount of tidal dissipation. This dissipation can drive spin states to synchronize, interiors to differentiate, volcanic centers to erupt, and orbits to evolve. All of these processes are driven by the tidal exchange of energy and similarly tidal dissipation should be a direct source of seismic energy release.

Energy dissipation in a spin-synchronous body due to reworking from orbital eccentricity can be defined as:

$$dE_T/dt = \left(\frac{k_2}{Q}\right) \left(\frac{21}{2} e^2\right) \left(\frac{G\, M_p^2\, n\, R^5}{a^6}\right) \quad (1)$$

where $k_2$ is the second order gravitational Love number of the body's response to the tide-raising potential, $Q$ is the quality factor describing the dissipation of energy per cycle within the body, $e$ is the orbital eccentricity, $M_p$ is the mass of the tide-raiser, $n$ is the mean motion of the body, $R$ is the body's radius, and $a$ is its orbital semi-major axis. This form of the energy dissipation equation assumes that eccentricity is not large (see e.g. Wisdom, 2008) and represents an average tidal dissipation over an orbital cycle. Thus, in order to accurately quantify the total energy dissipated, Eq. 1 needs to be integrated over a time period $t$ that represents full orbital cycles.

To resolve tidal dissipation rates on partial-orbit timescales, we employ the methods of Henning and Hurford (2014), whereby viscoelastic deformation for a body with an arbitrary number of laterally homogenous spherical shells is computed. As described in detail in Appendix A, this method allows us to determine heating as a complete function of sub-orbit time, as well as in three dimensions throughout the tidally active object. The basic parametric dependencies of this method are the same as shown in Eq. 1, particularly with regards to astrometric terms such as $e$, $a$, and $M_p$. The only difference is that the viscoelastic method allows us to determine dissipation not by pre-selecting estimates for the Love number $k_2$ and quality factor $Q$, but instead to compute equivalent effective $k_2$ and $Q$ values resulting from intrinsic material properties including the temperature, viscosity, density, and shear modulus of each layer. Using these methods, a more general expression for tidal dissipation can be found to be

$$dE_T/dt = \left(\frac{k_2}{Q}\right) \left(\frac{21}{2} e^2\right) \left(\frac{G M_p^2 n R^5}{a^6}\right) (1 - 0.143 \cos(2 n t)). \qquad (2)$$

This more general equation of tidal dissipation shows that the rate of energy dissipation fluctuates by $\sim\pm14\%$ throughout the orbital cycle.

Though the instantaneous tidal dissipation result is based on modeling tidal dissipation within a body comprising multiple layers with distinct material properties, the term $0.143 \cos(2nt)$ is not very sensitive to exact interior structure (see Appendix A). This means that while the amount of tidal dissipation does ultimately depend on the interior structure and the tidal response of the body, the change in dissipation rate does not. We have performed a sensitivity analysis of the coefficient of the cosine term in Eq. 2, and find it is fully independent of object sizes, tide-raiser distance, forcing periods, and overall tidal intensity, and varies only at the 4th significant digit based on layer structure alterations (with a systematic variation between a homogeneous and shell/asthenosphere-dominated tidal interior structure, see e.g., Beuthe 2013, Tyler et al. 2015). As described in these references, a highly degenerate set of interior models effectively reduces to a 1-dimensional spectrum of outcomes, with one extreme being a homogenous world, and the other a structure dominated by a thin shell or asthenosphere. All possible shell thicknesses (and viscosities) for Europa are naturally encompassed in this spectrum. From a homogeneous end member to a thin shell multi-layer model, the effect on the magnitude of the sub-orbit change in dissipation rate is <0.1% (0.1429-0.1430) and the timing in variability is not affected (See Appendix A for further discussion of this coefficient and its sensitivity). Therefore, for any body, regardless of its exact interior structure, we find that instantaneous dissipation rate would be well approximated by Eq. 2.

Using the tidal dissipation rate, the total energy dissipated in a given time period $T$ can be found as

$$E_T = \left(\frac{k_2}{Q}\right) \left(\frac{21}{2} e^2\right) \left(\frac{G M_p^2 n R^5}{a^6}\right) \int_0^T (1 - 0.143 \cos(2 n t)) \, dt. \quad (3)$$

Eq. 3 describes all of the energy lost to the interior of a body from eccentricity tides and in planetary applications it has been assumed that *all* of this energy is dissipated as heat within the interior. In reality, Eq. 3 represents the sum total of energy available for tidally driven processes, of which viscous heating is probably dominant. We propose here that a portion of this energy budget can be converted to seismic energy,

$$E_T = E_v + E_S. \quad (4)$$

That is the total energy, $E_T$, is partitioned into viscously dissipated energy that drives heating, $E_v$, and seismic energy, $E_s$. Furthermore, the partitioning into seismic energy depends on the efficiency of that partitioning, $E_s = \eta_0 E_T$.

Seismic energy, $E_s$ includes energy radiated as seismic waves, energy expended fracturing the rock, and energy dissipated a frictional heating during fault sliding. Seismic energy is proportional to the seismic moment, $M_0$, through the ratio of average stress on the fault, $\bar{\sigma}$, divided by the rigidity of the rocks, $\mu$, i.e. $E_s = \bar{\sigma} M_0/\mu$. This relationship allows us to ultimately tie the total moment released in seismic events to a fraction of $E_T$: $M_0 = \eta E_T$, where the constant of proportionality, $\eta = \eta_0 \mu/\bar{\sigma}$.

## 3 Modeling the size distribution of tidally-driven seismic events with a Gutenberg-Richter relationship

We assume that seismic activity in a tidally-active body follows the Gutenberg-Richter relationship (Gutenberg and Richter, 1944). This method has previously been used to estimate activity on Mars (Golombek et al., 1992) and Europa (Panning et al., 2018). The Gutenberg-Richter relationship quantifies the cumulative number of events $N$, equal to or greater than a particular seismic magnitude ($M_W$),

$$\log N(M_W) = a - bM_W. \tag{5}$$

In the relationship, the constants $a$ and $b$ are usually fit empirically to a catalog of known events. Using the fact that a seismic moment, $M_o$, can be related to the seismic magnitude by

$$\log M_o(M_W) = 1.5M_W + 9.1 \tag{6}$$

and following Golombek et al. (1992), the Gutenberg-Richter relationship can be recast as

$$N(M_o) = A\, M_o^{-2b/3} \tag{7}$$

where $a = \log A - 18.2b/3$. The constants $a$ or $A$, and $b$ can be fit empirically if a catalog of seismic events is available. However, in the absence of an observation catalog, a new method of estimating these values is needed.

As Golombek (1992) pointed out, the Gutenberg-Richter constants can be estimated by looking at the total moment released in the system. The total number of events $\delta N_T$ predicted for a given moment release between $M_o$ and $M_o + dM_o$ is related to the Gutenberg-Richter relationship by

$$\delta N_T(M_o) = -\frac{dN(M_o)}{dM_o}dM_o = \frac{2A\,b}{3}\,M_o^{-1-2b/3}dM_o \quad . \tag{8}$$

The moment released by all events within a certain size bin is estimated by $M_o\,\delta N_T(M_o)$. This estimate is most accurate when the size of the bin is fairly small such that the moment release from the representative event is not too different from the moment release from the largest event in the size bin. Finally, the total moment released by all the events described by the Gutenberg-Richter relationship is

$$\sum M_o = \int M_o\,\delta N_T(M_o) = \int_0^{M_C}\frac{2A\,b}{3}\,M_o^{-2b/3}\,dM_o = A\left(\frac{2b}{3-2b}\right)\,M_C^{1-\frac{2b}{3}} \tag{9}$$

where $M_C$ represents the moment of the largest event assumed possible. This result is valid for $b$-values less than 1.5, when the total seismic moment released is dominated by the less frequent, larger events. A similar method could be used to determine the number of events and total moment released for $b$ values greater than 1.5; the case where the total seismic moment released is dominated by the more frequent, smaller events. In that case, however, a lower limit to event size would need to be defined instead, which is non-trivial, as there is little observational evidence to indicate a lower limit to the self-similarity behavior of fault motions described by the Gutenberg-Richter relationship (e.g., Boettcher et al., 2009). Applications in this paper, therefore, are restricted to $b$-values < 1.5.

Finally, using the relationship between $A$ (or $a$), $\Sigma M_o$ and $M_C$, as shown in Eq. 9, the Gutenberg-Richter relationship for the cumulative number of seismic events with seismic magnitude $M_W$ or seismic moment $M_o$ can be written as

$$N(M_W) = \left(\frac{3-2b}{2b}\right) \sum M_o \ M_C^{\frac{2b}{3}-1} \ M_o (M_W)^{-2b/3} \quad \text{for } b < 1.5 \qquad (10)$$

and the total number of events as a function of seismic moment as

$$\delta N_T(M_o) = \left(\frac{3-2b}{3}\right) \sum M_o \ M_C^{\frac{2b}{3}-1} \ M_o^{-1-2b/3} dM_o. \qquad (11)$$

In order to predict a size distribution of seismic events using the Gutenberg-Richter relationship, there are three key parameters that must be specified. These parameters are: 1) the $b$-value of the slope of the distribution, 2) total seismic energy released, $\Sigma M_o$, and 3) a cutoff event size, $M_C$.

## 4 Constraints from the Earth-Moon System

The Moon is the only body other than the Earth where abundant seismic events have been recorded so far. While not all seismic activity within the Moon is tidally driven, deep and shallow quakes on the Moon have been associated with Lunar tides (Lammlein, 1977; Nakamura, 1978). Of the two types of moonquakes, deep quakes are smaller in seismic magnitude than the shallow quakes but more numerous (Lammlein, 1977). However, since shallow quakes dominate the moment release in the Lunar quake catalog, we focus on these events to link parameters for our model of tidally-driven seismic activity to tidal dissipation.

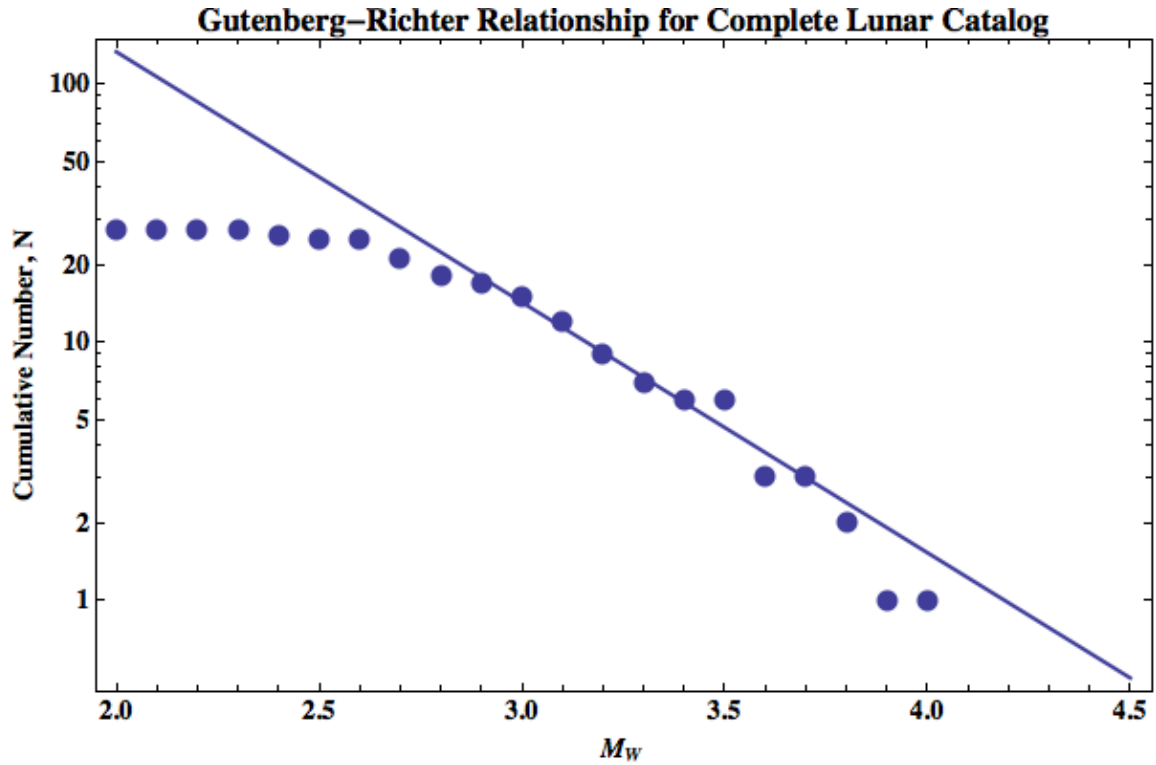

Figure 1. The Gutenberg-Richter relationship for the catalog of 28 shallow Lunar events. The catalog shows a fall-off in events smaller than seismic magnitude of 2.9 and the largest event has seismic magnitude ~4.0. The catalog of events with seismic magnitude ≥ 2.9 has a *b*-value of 0.97 with *a* = 4.06 (or *A* = 8.8 x 10^9).

The catalog of the 28 known lunar shallow quakes is too small to constrain well the *b* parameter. One analysis of the shallow seismic events yielded a low estimate for *b* of 0.5 (Nakamura, 1977), while another study of the data from the same population of events concluded, quite differently, that *b* can be as high as 1.78 (Lammlein et al., 1974). We suspect fits to the data yielded the different *b*-value estimates based on how lower magnitude events were treated and what portion of the catalog was used in the fits. Fig 1. shows the cumulative size distribution of the data points for this catalog of shallow events as reported by Oberst (1987) recast with the relationship, $\log M_o(M_W) = 1.5 M_W + 9.1$ shown in Eq. 6. We note that the cumulative number of events remains relatively constant at lower seismic magnitudes. We interpret that the data indicated the event catalog is incomplete for seismic magnitudes $M_W$<2.9. Either these small events are present and not detectable in the data from the seismic stations or the shallow events are a subset of a larger tidally-driven population and once deep moonquakes are included with this data the whole quake catalog would return to a linear distribution. Therefore, we use the events of seismic magnitude ≥2.9 to reevaluate the fit of the Gutenberg-Richter relationship to the catalog. We find the *b*-value that best represents the data is 0.97, close to the oft-adopted *b*-value of 1, and the *a* parameter that best fits the data is 4.06, which corresponds to a value of the *A* parameter of 8.8 x $10^9$.

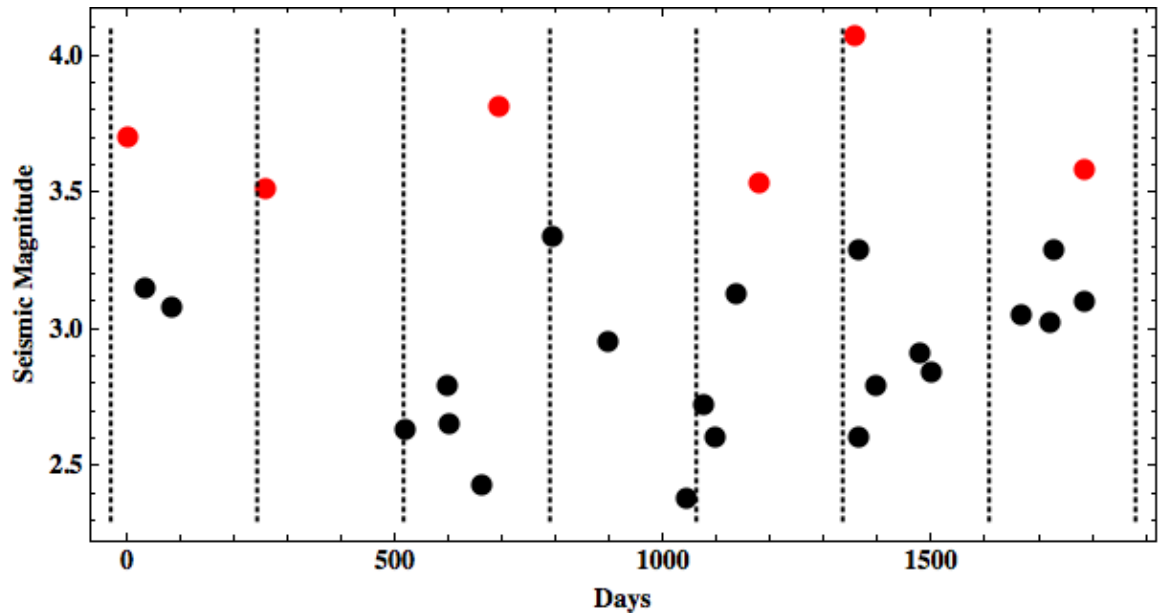

Figure 2. The catalog of shallow Lunar events is shown demarcated by dashed lines marking 10-cycle (273 day) periods. The seismic magnitude $M_W$ of each event is computed from the moment reported by Oberst (1987) using Eq. 6. Events with seismic magnitude ≥3.5 are shown in red.

Because moment release in a population of earthquakes characterized by a Gutenberg-Richter relationship with *b*-values near 1 is dominated by the largest events, seismicity rates predicted from the total moment release are strongly dependent on the choice of maximum event size. If the maximum event size is large relative to the cumulative moment release during a relevant time period such as a tidal cycle, moment release will be dominated by relatively infrequent large events and most tidal cycles will experience smaller moment release than that expected from the long-term average. On the other hand, if the moment of the maximum event is comparable to the cumulative moment release over a tidal cycle, cumulative moment release will match the long-term average over a smaller number of tidal cycles. Therefore, we can interpret the maximum event size as controlling whether moment release will be relatively consistent across tidal cycles, or dominated by rare events. In order to constrain this, we examine the catalog of 28 shallow lunar seismic events that occur over a period of ~65 months recorded from 1971 to 1976. While Oberst (1987) and others tend to look at lunar data on yearly periods, this timescale is somewhat arbitrary. In our framework for tidally-driven seismic activity, we need to determine on what timescale the seismic moment (our proxy for seismic energy) added to the system balances the seismic moment released from all the seismic events. For the Lunar seismic record, large events do not appear occur on a monthly basis indicating this timescale is larger than one tidal cycle or 27.3 days. The catalog of shallow seismic events shows there are 6 events with seismic magnitudes ≥3.5, which seem to occur fairly regularly throughout the ~65-month record (Fig. 2). We take this as evidence that the timescale on which approximate energy balance is achieved is at least ~10 orbital cycles or ~273 days. The maximum timescale for achieving approximately energy balance supported by the observational record is ~65 orbital cycles but since large events happen frequently with no single biggest event relieving most of the seismic momentum, we favor the 10-cycle timescale.

Observations of the evolution of the Moon's pole of rotation by laser ranging suggests strong dissipation within the Moon and constrains the value of $k_2/Q$ to be ~0.0012 (Williams et al., 2001), yielding a tidal dissipation rate of 1.18 GW of energy. This rate of tidal dissipation results in ~2.8 x $10^{16}$ J of energy dissipated due to eccentricity tides inside the Moon in 10 orbital cycles. For the Moon, there is also dissipation from obliquity tides, which increases the total dissipation to ~5x$10^{16}$ J. To define the factor that describes the conversion of tidally dissipated energy to total seismic moment released, we examine the total moment released as reported by Oberst (1987). Over 65 tidal cycles, ~4x$10^{15}$ Nm of moment (seismic magnitude equivalent of 4.33) is released by the largest events. But using Eq. 9 with parameters fit for the Gutenberg-Richter relationship noted above and a $M_C$ = 3.55x$10^{15}$ Nm ($M_W$ = 4.3), we can estimate that the total moment release is likely closer to ~5x$10^{15}$ Nm ($M_W$ = 4.4), once smaller events are taken into account. Note that $M_C$ is approximately 70% to the total estimated moment released. Given that ~3x$10^{17}$ J of energy in total is dissipated over ~65 cycles the conversion factor between energy dissipated and moment released is ~0.017 Nm/J. Applying this factor to the energy release in 10-orbital cycles, ~5x$10^{16}$ J of energy dissipated, results in an average moment release of ~8x$10^{14}$ Nm ($M_W$ = 3.9). If on the 10-orbital cycle timescale the ratio of $M_C$ to the total moment release is ~70% and the largest events have a moment release of ~4.9x$10^{14}$ Nm ($M_W$ = 3.7), which is consistent with the observational record. While we assumed a balance in the moment built up and released on the 10-orbital cycle timescale, this is a long-term average and since large events are stochastic, it is expected that there can be variations in moment release and the size of large events from cycle to cycle as seen in the Lunar data.

In summary from the lunar seismic record we find:

- The timescale for moment balance is at a minimum 10 orbital cycles.
- The total moment released, $\Sigma M_o$, by Lunar seismic events on average over 10 orbital cycles is $\sim 8 \times 10^{14}$ Nm ($M_W = 3.9$).
- The conversion factor $\eta$ between energy dissipated by tidal dissipation from both Lunar orbital eccentricity and obliquity, and the total moment released is $\sim 0.017$.
- The cutoff event size, $M_C$, represents $\sim 70\%$ of total moment released $\Sigma M_o$.

## 5. Implications for Seismic Activity on Tidally Active Worlds

With a model of tidally driven seismic activity and constraints based on the Earth-Moon system, the tidally driven seismic activity can be simulated for other tidally active worlds and the largest seismic event expected on different tidally active worlds can be predicted. It follows from our model that the largest event predicted is proportional to the total seismic moment released by

$$M_C = f \sum M_o \tag{12}$$

where $f$ is the fraction of the total moment release captured in the largest event. This can further be expanded to

$$M_C = f \, \eta \left(\frac{k_2}{Q}\right) \left(\frac{21}{2} \, e^2\right) \left(\frac{G \, M_p^2 \, n \, R^5}{a^6}\right) T \tag{13}$$

where $\eta$ represents the conversion factor between total energy dissipated and total seismic moment released while $T$ is the characteristic timescale for moment balance.

The largest seismic events for a number of tidally active worlds obtained from equation 13 are shown in Table 1. Predictions reported here are based on assuming that 1) $f = 0.7$, 2) $\eta = 0.017$, and 3) $T = 10$ orbital cycles.

Table 1. Predictions of the largest seismic events on various tidally active worlds. In the table the first 6 columns ($M_P$, Period, R, a, e and $k_2$/Q) are the parameters needed to evaluate the energy dissipation. Column

$E_T$ is the total energy dissipated over 10 orbital cycles. Based on the total energy dissipated we calculate the total momentum released $\Sigma M_o$ and present the result in terms of [Nm] and the equivalent seismic magnitude $M_W$. Finally, we estimate the largest predicted seismic event $M_C$ and again present the result in terms of [Nm] and the equivalent seismic magnitude $M_W$.

| World | $M_P$ [kg] | Period [days] | R [km] | a [km] | e | $k_2/Q$ | $E_T$ [J] | $\Sigma M_o$ [Nm] | $M_W$ | $M_C$ [Nm] | $M_W$ |
|---|---|---|---|---|---|---|---|---|---|---|---|
| Io | $1.9\times10^{27}$ | 1.769 | 1821.6 | 421700 | 0.0041 | $0.015^a$ | $1.43\times10^{20}$ | $2.3\times10^{18}$ | 6.2 | $1.7\times10^{18}$ | 6.1 |
| Europa | $1.9\times10^{27}$ | 3.551 | 1560.8 | 670900 | 0.01 | $0.0054^b$ | $8.7\times10^{18}$ | $1.5\times10^{17}$ | 5.4 | $1.\times10^{17}$ | 5.3 |
| Titan | $5.68\times10^{26}$ | 15.945 | 2575 | $1.22\times10^6$ | 0.0288 | $0.004^c$ | $1.6\times10^{18}$ | $2.7\times10^{16}$ | 4.9 | $1.9\times10^{16}$ | 4.8 |
| Moon | $5.97\times10^{24}$ | 27.3 | 1737.1 | 384399 | 0.055 | $0.0012^d$ | $5.\times10^{16}$ | $8.\times10^{14}$ | 3.9 | $4.9\times10^{14}$ | 3.7 |
| Enceladus | $5.68\times10^{26}$ | 1.37 | 252 | 237948 | 0.0047 | $0.0036^b$ | $6.3\times10^{15}$ | $1.\times10^{14}$ | 3.3 | $7.5\times10^{13}$ | 3.2 |
| TRAPPIST 1b$^f$ | $1.59\times10^{29}$ | 1.51 | 6919 | $1.66\times10^6$ | 0.081 | $0.022^e$ | $1.21\times10^{26}$ | $2.06\times10^{24}$ | 10.1 | $1.44\times10^{24}$ | 10. |
| TRAPPIST 1c$^f$ | $1.59\times10^{29}$ | 2.42 | 6728 | $2.28\times10^6$ | 0.014 | $0.022^e$ | $4.75\times10^{23}$ | $8.1\times10^{21}$ | 8.5 | $5.6\times10^{21}$ | 8.4 |
| TRAPPIST 1e$^f$ | $1.59\times10^{29}$ | 6.1 | 5849 | $4.22\times10^6$ | 0.007 | $0.022^e$ | $1.46\times10^{21}$ | $2.5\times10^{19}$ | 6.9 | $1.7\times10^{19}$ | 6.8 |
| Kepler 20e$^g$ | $1.81\times10^{30}$ | 6.1 | 5530 | $7.58\times10^6$ | 0.14 | $0.022^e$ | $1.7\times10^{24}$ | $2.9\times10^{22}$ | 8.9 | $2.\times10^{22}$ | 8.8 |
| Kepler 20f$^h$ | $1.81\times10^{30}$ | 19.58 | 6562 | $1.65\times10^7$ | 0.16 | $0.022^e$ | $4.89\times10^{22}$ | $8.3\times10^{20}$ | 7.9 | $5.8\times10^{20}$ | 7.8 |
| 55 Cancri e$^i$ | $1.91\times10^{30}$ | 49.41 | 12232 | $2.31\times10^6$ | 0.05 | $0.022^e$ | $1.59\times10^{28}$ | $2.7\times10^{26}$ | 11.6 | $1.89\times10^{26}$ | 11.5 |
| HD 219134b$^j$ | $1.58\times10^{30}$ | 3.094 | 10232 | $5.71\times10^6$ | 0.065 | $0.022^e$ | $3.28\times10^{25}$ | $5.6\times10^{23}$ | 9.8 | $3.9\times10^{23}$ | 9.7 |
| Tau Ceti b$^k$ | $1.56\times10^{30}$ | 13.96 | 8282 | $1.57\times10^7$ | 0.16 | $0.022^e$ | $1.56\times10^{23}$ | $2.7\times10^{21}$ | 8.2 | $1.9\times10^{21}$ | 8.1 |

[a]Lainey et al. (2009); [b]Vance et al. (2007); [c] Vance et al. (2018); [d]Williams et al. (2001); [e] Earth analog value using $k_2 = 0.29$ (Kozai, 1968) and Q = 13 (Goldreich and Soter, 1966); [f]Wang et al. (2017); [g]Buchhave et al. (2016); [h]Buchhave et al. (2016); [i]Demory et al. (2016); [j]Motalebi et al. (2015); [k]Teixeira et al. (2009)

Most of the satellites in the Solar System examined here, should experience seismic events of larger seismic magnitude than the largest Lunar seismic events. And because these satellites have shorter orbital periods, these large events will occur more frequently than on the Moon. The smallest Solar System satellite studied, Enceladus, should have seismic events that are comparable in seismic magnitude to Lunar events, but again these will happen more frequently, about twice a month compared to once every 10 months on the Moon.

We predict that all of the exoplanet bodies studied should experience large seismic events. These worlds are very close to their host stars and dissipate large amounts of energy, if it is assumed they are at least as dissipative as the Earth (e.g. $k_2/Q = 0.022$). And moreover, because of their short periods, these worlds should experience large seismic events quite frequently.

With the parameters of total seismic energy released, $\Sigma M_o$, and a cutoff event size, $M_C$ as defined in Table 1, idealized size frequency distributions for seismic activity on these worlds can be defined using the Gutenberg-Richter relationship for any *b*-value less than 1.5. For a *b*-value of 1, many of these worlds will experience numerous smaller events during a 10-orbit period.

## 6. Application to Europa

Tidal dissipation on Europa allows a subsurface ocean to persist. Associated tidal stresses have fractured its surface, which is cross-cut by numerous faults, suggesting that tides can drive faulting and give rise to seismic events across its surface. Therefore, Europa is likely currently seismically active; it has been proposed that this natural seismic activity will excite seismic waves that can be detected by seismometers to explore Europa's interior (Kovach and Chyba, 2001; Lee et al., 2003; Lee et al., 2005; Panning et al., 2006). Of all the tidally active worlds listed in Table 1 from the Solar System, we focus on Europa as a case study to explore further how the tidally driven seismic model can make more predictions on seismic activity. The future possibility of a lander mission to explore Europa makes it especially important to estimate its seismic activity.

In order to constrain seismic activity for Europa, three critical parameters need to be estimated: the *b*-value of the slope of the distribution, total seismic energy released, $\Sigma M_o$, and a cutoff event size, $M_C$. Table 1 provides estimates for $\Sigma M_o$ and $M_C$ based off our Lunar scaling, leaving only the *b*-value unconstrained. Again, the Moon analysis of event catalogs shows a wide range of possible *b* values from 0.5-1.78 (Nakamura, 1977; Lammlein et al., 1974) and the analysis here supports a *b*-value of ~1. Earth event catalogs generally have values that range from ~0.7 to ~1.3 (Frohlich and Davis, 1993); hence, it seems that for rock, *b*-values around 1 would be applicable. For terrestrial icequakes, studies show a bimodal distribution: some studies yield a *b*-value near 1, while others yield higher values near 2 (Podolskiy and Walter, 2016). Therefore, we assume a b-value of 1 to be at least plausible for Europa,.

With the seismic model parameters constrained, the Gutenberg-Richter equations can be used to produce the idealize size distribution for Europa seismic activity. Moreover, the relationship can also be used to estimate of the probability of an event occurring in a given time period, enabling us to build a stochastic models of seismic activity. The probability of an event occurring of any magnitude is roughly the number of events of that magnitude forming per second when the rate of formation is <<1/second. The probability of an event of magnitude $M_W$, in one second, can then be written as:

$$P(M_W) \approx \frac{\delta N_T(M_o)}{T} \approx \left(\frac{3-2b}{3T}\right) \sum M_o \; M_C^{\frac{2b}{3}-1} \; M_o^{-1-2b/3} dM_o. \qquad (14)$$

This probability would be valid if the rate of energy dissipation within a tidally-active body were uniform with time. This is effectively the approach used by Panning et al. (2018), in which event probability was assumed to be uniform in time and space. However, since the rate of energy dissipation is not constant in time, the tidal dissipation equation (Eq. 1) details the average rate of dissipation, which is most accurate for timescales that are multiples of the orbital period.

This equation continues to average over the full object volumetrically, while eliminating the averaging with respect to time. For objects such as Europa, the dissipation (in solid layers) outside of the ice shell is negligible, and therefore this equation also effectively represents the near-surface dissipation rate. We do not address any dissipation occurring in fluid layers (Tyler, 2008). While the total number of seismic events depends on the total seismic energy released, the instantaneous probability of events occurring can be expected to be proportional to the instantaneous rate of energy dissipation. This expectation is reasonable because the rate of energy dissipation will be proportional to the rate of change of all terms describing the stress and strain tensors throughout the object and over the course of 10 orbital cycles the total moment balance is still achieved. Thus, when the rate of energy dissipation is higher, the likelihood of an event occurring should increase, and when the rate of energy dissipation is lower the likelihood should decrease. This change in production rate modifies the event probability to correspond to the change in the rate of energy dissipation,

$$P(M_W, t) \approx \left(\frac{3-2b}{3T}\right) \sum M_o \; M_C^{\frac{2b}{3}-1} \; M_o^{-1-2b/3} dM_o (1 - 0.143 \; \cos(2 \; n \; t \;)) \quad . \qquad (15)$$

The effect of the tidal dissipation rate and the triggering of seismic events is estimated to cause a 14% increase in earthquake occurrence probability at one- and three-quarters of an orbit with a corresponding 14% decrease in formation probability at pericenter and apocenter. This occurs because it is at these times that the librational tide, rather than the radial tide, is at a maximum. The librational tide occurs due to the fact that for a spin-synchronous body in an eccentric orbit, a vector normal fixed to the sub-perturber point at pericenter will not remain pointed to the perturber itself at other points within the orbit (but in fact points to the orbit's empty focus). The radial tide by contrast is due only to sub-orbit changes in the perturber-target distance. However, the total number of events predicted over the course of a full orbit is not affected by this variation, and any variation from orbit to orbit would be caused by the fact that events are stochastic.

Using this stochastic seismic model, one realization of seismic activity over 10 Europan orbits can be constructed (Fig. 3-*top*). As seen in Figure 3-*middle*, the size distribution of events in the synthetic catalog follows the idealized Gutenberg-Richter relationship using the parameters in Table 1. Slight variations are seen as a result of the stochastic nature of the model and become more prominent at higher event magnitudes since here the variability due to the small sample size are heightened. When the events are binned by the orbital phase (Fig. 3-*bottom*), seismic activity at the quarter orbit periods is seen to be enhanced compared to activity at pericenter and apocenter.

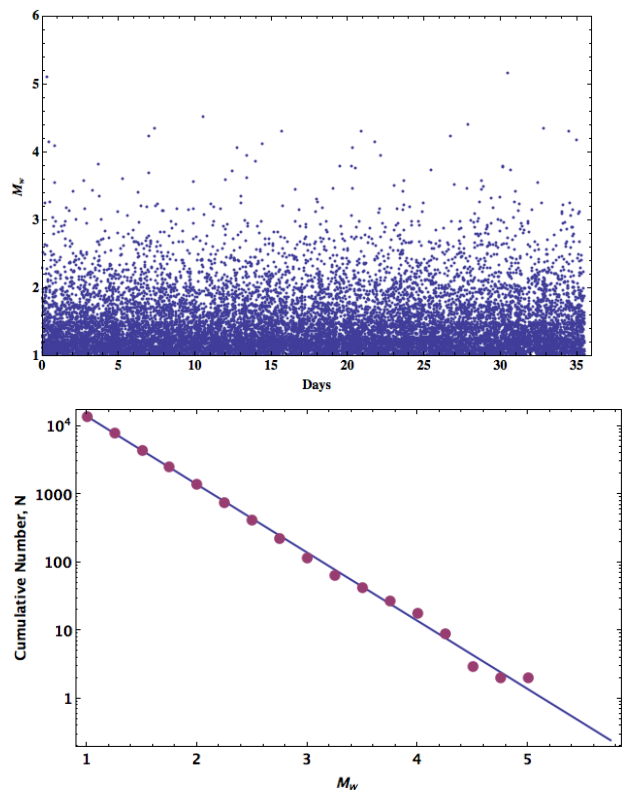

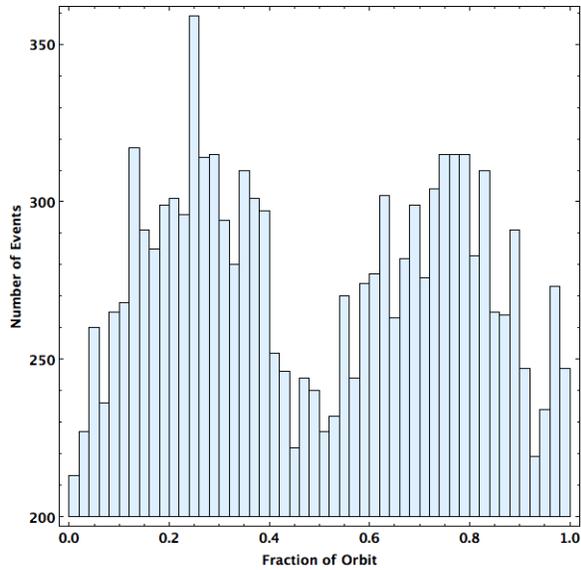

Figure 3. *Top:* A seismic catalog of events produced over a 10-orbit cycle (35.51 days) is shown with event magnitude plotted as a function of time. *Middle:* The size distribution in this one realization of a stochastically determined seismic catalog of events (dots) follows the idealized case represented by the Gutenberg-Richter relationship with parameters form Table 1 and a *b*-value of 1 (line). *Bottom:* The event catalog plotted with the number of events binned by orbital phase from pericenter shows the effect of the instantaneous dissipation rate of event formation with more events occurring at one-quarter and three-quarters of an orbit and fewer events at pericenter and apocenter.

If the rate of tidal dissipation were uniform within the interior of the body, there would be no preference for where a seismic event would occur. Hence, without more information about how seismic activity may be concentrated in the body, the event could be assigned to a random position in latitude and longitude. But just as the instantaneous tidal dissipation rate is likely to introduce variation in the timing of seismic events, the spatial variations in tidal dissipation rate may drive variations in the likely locations of seismic activity.

Indeed, the rate of tidal dissipation is not uniform across the surface, and even the pattern of spatial heterogeneity changes throughout the orbit (Fig. 4). In Figure 4 the rate of tidal dissipation on Europa is calculated with a multilayer modeling approach (Henning and Hurford, 2014). For each point on the surface, the rate of tidal dissipation radially beneath that point is integrated such that the tidal dissipation pattern plotted reflects total variations in heating within Europa. This pattern is sensitive to the interior model assumed; in Figure 4, it is computed for a five layer model of Europa, whose parameters are specified in Table 2. For Europa, heating solutions are insensitive to material parameters below the ice shell, yet quite sensitive to ice shell thickness and viscosity. For computational stability reasons, we include small offsets in the densities of ice and water layers, and assume that the ocean (non-inertial) response is well-approximated by a very weak solid layer with shear modulus 1 x $10^5$ Pa, and viscosity of $10^5$ Pa s (Henning & Hurford 2014).

Table 2. Model parameters for a five layer Europa.

| | Thickness (km) | Density (kg/m3) | Shear Modulus (Pa) | Viscosity (Pa s) |
|---|---|---|---|---|
| Iron Core | 391 | 8000 | $6.5 \times 10^{11}$ | $1 \times 10^{21}$ |
| Silicate Mantle | 1044.8 | 3500 | $6.5 \times 10^{10}$ | $1 \times 10^{20}$ |
| Ocean | 100 | 1000 | negligible | negligible |
| Ductile Ice | 20 | 999 | $3.49 \times 10^9$ | $1 \times 10^{14}$ |
| Brittle Ice | 5 | 998 | $3.49 \times 10^9$ | $1 \times 10^{21}$ |

Locations (latitude, longitude) are assigned to each event predicted in the stochastic catalog based on a likelihood value, $L$(lat,lon,t), based on the rate of tidal dissipation at that location at that moment in the orbit and the minimum and maximum rates of tidal dissipation experienced anywhere throughout the orbit. More specifically,

$$L(\text{lat,lon,t}) = \frac{\frac{dE}{dt}|_{\text{lat,lon}} - \text{Min}(\frac{dE}{dt})}{\text{Max}(\frac{dE}{dt}) - \text{Min}(\frac{dE}{dt})} \qquad . \qquad (16)$$

The randomly generated surface position is then tested against this likelihood value to see if it is assigned to the seismic event. If it passes the test, it becomes the seismic location for that event. However, if it fails, a new random latitude and longitude is generated and tested until a valid location is determined. This method allows any point on the surface to be eligible for seismic activity but biases the results to locations of higher rates of tidal dissipation.

In addition to illustrating the pattern in the rate of tidal dissipation within the orbit, Figure 4 also shows the locations of seismic events predicted at each of these timesteps, with each covering 5% of the orbital period. As expected, seismic events are more numerous in regions where higher rates of tidal dissipation are exhibited (warmer colors in Figure 4). At apocenter and pericenter, when the overall seismicity rate diminishes by ~14%, the seismic events are preferentially clustered along longitudes corresponding to the center of Europa's leading and trailing hemispheres (90° and 270°). Also, the region near the sub-/anti-Jupiter point always experiences lower rates of tidal dissipation; as a result, it generates relatively few seismic events. This may be understood by thinking not in terms of simple radial deformation, but by the fact that tidal dissipation is primarily caused by shearing, (stress and strain tensor term blends of $\sigma_{ij}$, and $\epsilon_{ij}$ where $i \neq j$, see Appendix Eqn A5) and shearing is often greatest along the axes orthogonal to the axis of tidal symmetry.

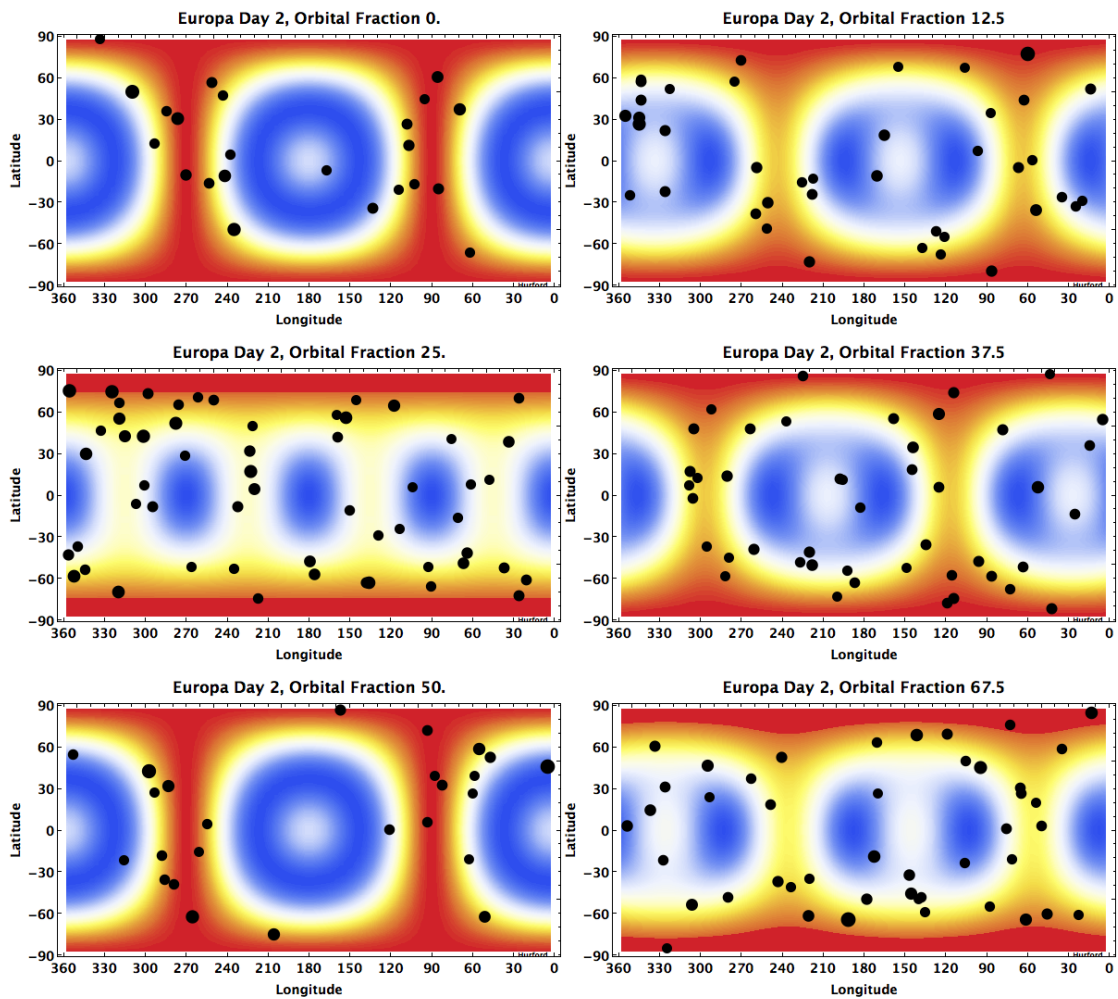

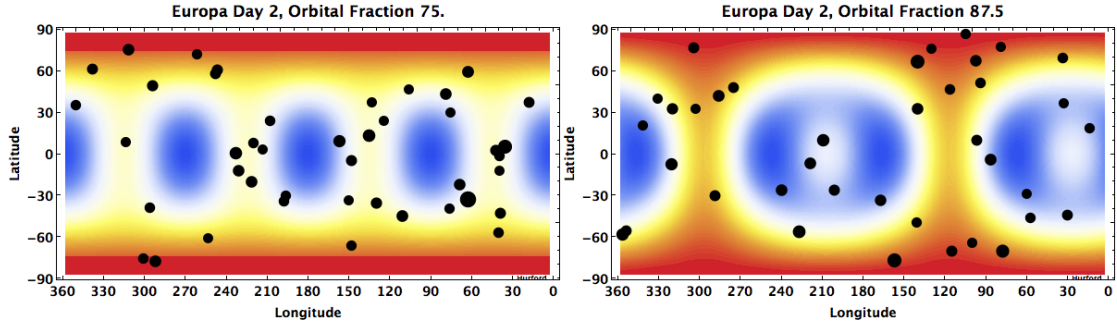

Figure 4. Seismic events predicted during Europa's orbit are shown at intervals of 1/8 of an orbit. In all cases, the contours show the spatial distribution of the rate of tidal dissipation at each timestep in relation to the predicted locations of seismic events. Warmer colors represent relatively higher rates of tidal dissipation and cooler colors relatively lower rates. The sum of all timesteps leads to the typical polar-dominated tidal heating pattern predicted for Europa (see e.g., Ojakangas & Stevenson, 1989 Figs 1−5). Seismic events predicted to form during these timesteps, each $1/20^{th}$ of an orbit in length, are plotted with respect to the rates of tidal dissipation. Points used to represent the locations of seismic activity are scaled in size based on the magnitude of the event. (See online material for seismic activity over the full 10-cycle period of 35.51 days.)

# 7 Discussion

Our model of tidally-driven seismic activity allows us to generalize the conversion of tidally dissipated energy into seismic activity on any body experiencing tidal dissipation, with the assumption that the scaling parameter, $\eta$, does not vary greatly as a result of internal structure. Many satellites in our Solar System should experience as much or more seismic activity as the Moon since tidal dissipation is greater on many of them. Indeed, the presence of significant fractures and evidence for recent tectonic and/or volcanic activity exists on most of the Solar System bodies modeled here. In exoplanet systems, tidal dissipation may play an even greater role, and close-in exoplanet systems should dissipate considerable energy capable of driving even higher seismicity rates. In our model, we predict that tidal dissipation could produce quakes on these bodies even larger than those experienced on Earth. If the strength of rock limits the maximum seismic release equivalent to a magnitude 9.5 earthquake (the maximum recorded event on Earth), then these largest events predicted by our model would not be physical. Instead, a larger number of smaller events would be needed to release all the seismic moment built up from tidal dissipation. In essence, this will require that the *b*-value be >1 and that the seismic moment release by smaller events plays a larger role in the total moment release. And just like our example for Europa, the location and timing of seismic events will vary on these bodies, but without further information it is too premature to conjecture about this variation.

However, for Solar System bodies, we can start to study the non-uniform nature of seismic activity and our model leads to the prediction that the rate of seismic activity is expected to vary throughout an orbital cycle. As shown in the case study for Europa, an increase in seismic activity is expected at one quarter and three quarters of an orbit and a similar reduction in activity is predicted at pericenter and apocenter. Lammlein (1977) plotted the number of Lunar quakes over a roughly two-year period from April 22, 1972 to May 21, 1974. The Lunar data does indeed show that lunar quakes are not randomly distributed over the Lunar orbital cycle but there are peaks in activity at a period of 13.6 days (half the Lunar cycle). For many cycles the peaks in activity do appear to line up with the one-quarter orbit and three-quarter orbit time frames, but the peak does sometimes also seem to line up with pericenter or apocenter. Previous studies have tried to correlate these peaks with tidal activity as given by the latitudinal or longitudinal libration of the Lunar tidal bulge (Lammlein, 1977). Therefore, the Lunar record is consistent with our model's prediction of tidal cycling of seismic activity. Furthermore, our model predicts that this activity should peak at the quarter orbit periods, which cannot be tested, because, as Lammlein points out, the Lunar record is probably incomplete. For example, Lammlein only counted events if they were recorded by multiple seismic stations. Moreover, the stations do not record the complete Lunar seismic record and proximity to each other captures just the activity local to all of them, with a bias to the near side of the moon. Still the changes in Lunar seismicity do fit our model and future Lunar seismic studies can further catalog seismic activity and test this prediction of the tidally-driven seismic model.

The static images in Figure 5 do not give a full picture of the variation of the rate of tidal heating or seismic activity. Animating the results for Europa over the course of an entire orbit (see Supplemental Online Information) shows that regions with high tidal dissipation rates migrate eastward throughout the orbit. While these zones traverse the surface, seismic activity migrates with them. This implies that the ambient seismic background noise may fluctuate with regional variations in seismic activity. Panning et al. (2018) simulated ambient noise for a Europa model in which seismic events were randomly located and produced at a constant rate throughout the orbit. However, at any one location the ambient noise may rise and fall as the seismic events sweep past the region.

Noting that the pattern of dissipation exhibits a generally eastward (in the direction of spin) migration relative to the surface (for an object with both prograde revolution and rotation), may at first seem unexpected given that the total tidal potential, the function that describes the tidal bulge in time, merely librates and fluctuates in magnitude, with the bulge generally centered near the sub-/anti-perturber points. The eastward drift however occurs because tides depend not on the *total* tidal potential, but on its time varying component. After time-independent terms are subtracted, the tidal potential exhibits the eastward motion shown, and lends this behavior to the dissipation map (see Appendix Eqs A2 and A3).

An implication of the spatial variations in seismic event occurrence predicted by our model is that solely characterizing the spatial distribution of seismic activity may also serve to constrain the interior of these bodies, complementing constraints from more traditional analysis of body and surface wave travel times and waveforms. Event locations are not only sensitive to interior structure, but also to how a body dissipates tidal energy. Even if a full characterization of seismic waves related to the event is not fully captured, knowing when and where an event has occurred can help to constrain the internal structure of a body.

A final implication of the spatial heterogeneity of seismic occurrence rates predicted by our model is that certain locations might be preferred in order to detect seismic activity on a tidally-active body. Figure 5 shows the spatial distribution of the seismic catalog used for the Europa case example (Figs. 3&4). Over the course of 10 orbital cycles the pattern of seismic events is not uniformly distributed across the surface of Europa. As a result, unless an instrument can detect events at longer ranges, seismometers in the sub-/anti- primary regions may not be as effective in detecting seismic events (Fig. 5-*top*). Similarly, a seismometer in the regions of the trailing and leading hemisphere might take advantage of the clustering of events during the orbital variation of seismic activity. However, as these results are sensitive to models of interior structure, a more careful exploration of possible seismic activity must be undertaken before any conclusions can be drawn.

While our test case catalog represents one realization of seismic activity on Europa, generating 100 different cases allows statistical analysis of seismic activity to be completed. Using 100 simulations of tidally-driven seismicity over 10 orbits, we can produce synthetic hazard maps of the surface of Europa. To produce these hazard maps we had to compute the accelerations experienced at any point on the surface due to seismic waves excited by these events. We use the spectral-element-method based code AxiSEM (Nissen-Meyer at al. 2014) to calculate seismic accelerations at frequencies up to 1 Hz. We assume a normal faulting source (e.g., Nimmo and Schenk, 2006), and calculate waveforms at seismic stations between 0 and 180 degrees epicentral distance. At each station, we measure the Rayleigh wave acceleration on the vertical component in order to determine how its amplitude decays with distance due to both geometrical spreading and attenuation. Our simulations use physically self-consistent models of Europa's interior seismic velocity and density structure (Cammarano et al. 2006), and assume a 20-km thick ice shell. The seismic shear quality factor $Q\mu$ is an important parameter for determining how Rayleigh wave acceleration decays with distance, yet $Q\mu$ in Europa's ice shell is not well constrained. A typical approach is to scale $Q\mu$ using a homologous melting temperature (e.g., Cammarano et al. 2006, Panning et al. 2018), however this approach depends on an assumed temperature profile and is often scaled arbitrarily to create a range of acceptable $Q\mu$ values. Additionally, scattering a near surface regolith layer can further attenuate waves and is not accounted for by intrinsic attenuation. Here, we choose to fix $Q\mu = 200$ within the ice shell. Europa attenuation models based on homologous temperature scaling typically have low attenuation in the low temperature regions near the surface, with $Q\mu > 1000$ (e.g., Panning et al. 2018). Thus, our modeling provides conservative estimates of the spatial decay of high frequency accelerations.

To construct the hazard maps, we compute the acceleration experienced at different points on the surface due to the surface waves produced by each event in each of the 100 event catalogs, assuming that the seismic depth's are sufficiently shallow to negligible affect simulated accelerations. We record the number of times a 100 ng acceleration (measured at 1 Hz) is exceeded at that point for each of the event catalogs and average that number over the 100 catalogs to produce a sample seismic hazard map (Fig. 5-*bottom*). For the case presented in this paper, we find that 100 ng acceleration from surface waves happen between 60 and 120 times in a 35.51 day period on Europa with the greatest activity near the poles and the lowest at the sub-/anti- Jupiter regions. This activity would be equally distributed throughout the 10 orbital cycles meaning that 100 ng surface waves would be expected to occur 6 to 12 times per orbit. Future work can extend to higher frequency wave propagation simulations, which should allow us to account for body waves accelerations, regions close to the epicenter, or on the effects of seismic event depth.

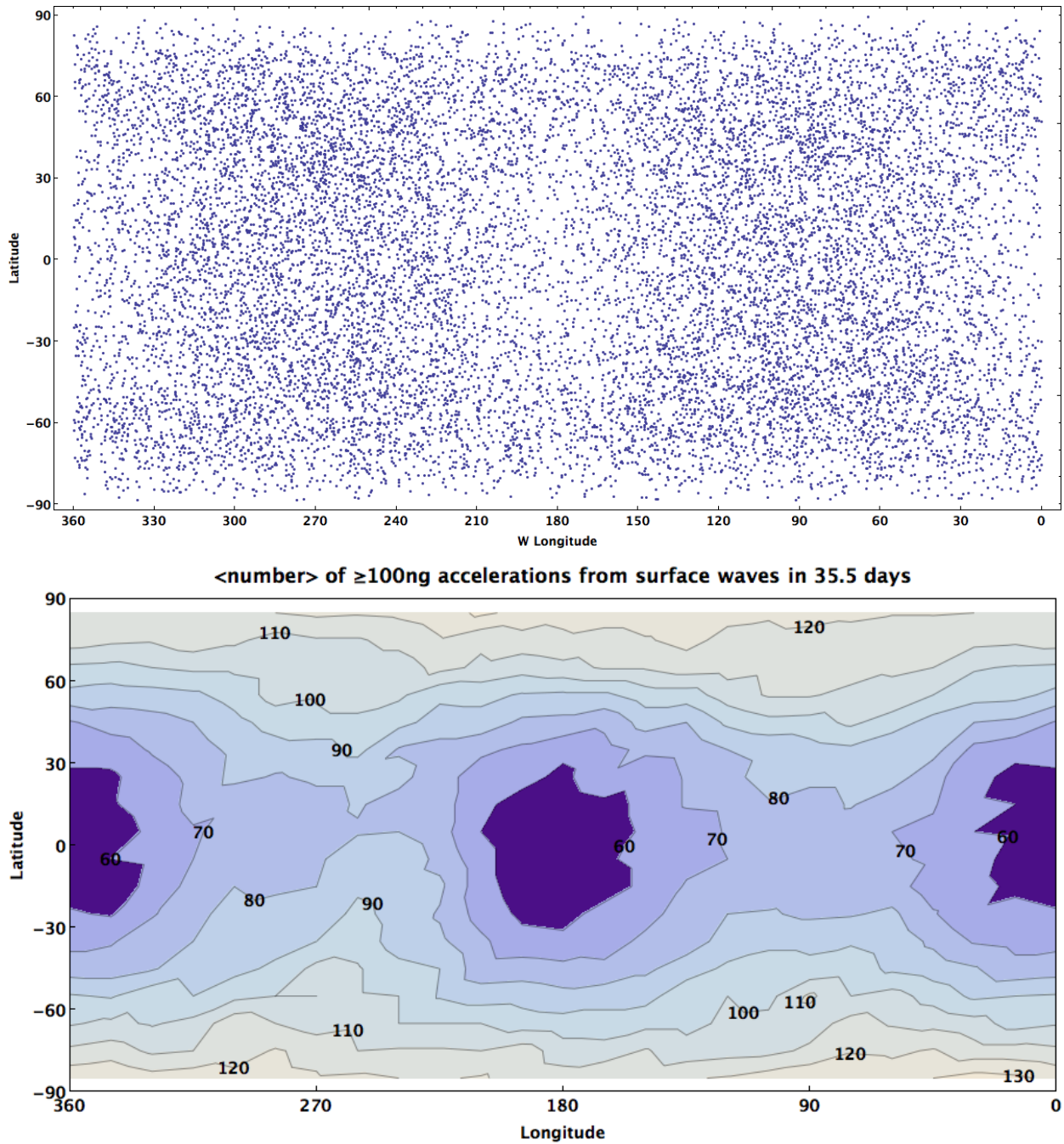

Figure 5. *Top* The seismic catalog of events produced over a 10 orbit cycle (35.51 days) is plotted by latitude and longitude of the event, showing the heterogeneous distribution across Europa's surface.

*Bottom* Using 100 uniquely generated seismic catalogs, a hazard map of the Europa surface shows the spatial variations of the average number of surface wave events ≥ 100 ng at 1 Hz accelerations.

## 8 Conclusions

Here we present an approach for directly linking seismic activity to tidal dissipation within tidally-active bodies. We find that only a small portion of the total energy dissipated within planet is likely converted into seismic energy as expressed by the total seismic moment released. Based off of the relationship between tidal dissipation and seismic activity on the Moon, the efficiency in converting tidal energy to seismic moment appears to be ~1.7%. Using this efficiency factor, we predict the seismic activity on a number of different Solar System bodies and exoplanets. The greater the energy dissipated within a body, the more seismic activity is expected. Short-period silicate/ice-surface exoplanets, where strong orbital perturbations exist to maintain nonzero eccentricity, of which TRAPPIST-1b is an archetypal example, are likely very seismically active.

Besides predicting the total seismic activity, variations in the rate of energy dissipation can lead to changes in the rate of seismic event occurrence. Regardless of the interior structure of a body, the variation in seismic triggering should result in ~14% fewer events at pericenter and apocenter and ~14% more events at a quarter and three-quarters of an orbit.

Moreover, the spatial variation in the rate of tidal dissipation should also be evident in the seismic record. However, as these patterns are dependent on a body's internal structure, they must be explored on a case by case basis. Here we have presented only one model of how Europa's internal structure might affect the spatial distribution of expected seismic activity.

## Appendix A. Tidal Heating Calculations

We resolve the tidal heating of an object in a non-orbit averaged manner by adapting the methods of Henning & Hurford (2014) for a multilayer viscoelastic object that were performed using orbit-averaging. This allows us to obtain both the global heating rate as a function of time throughout an orbit, as well to resolve the heating as a full 4D function in both time and 3 spherical coordinates.

This technique is based on the propagator matrix method (Love 1927; Alterman et al. 1959; Takeuchi et al. 1962; Sabadini and Vermeersen 2004), which has been used extensively for both Solar System applications, as well as extrasolar planet applications. In this method, a layer structure in the object under tidal flexure is first prescribed. For our Europa model, we use structure with 5 unique materials, and then resolve these layers further into sublayers to resolve how tidal heating varies with depth, particularly throughout the ice shell. See Table 2 for layer properties. Values were varied severely from those in Table 2 to verify insensitivity of the cosine coefficient in Eqn 3, to input parameters (This included but was not limited to, variations in densities, to a total body mass from ~1/10th to ~10x that of Europa, ice thickness variation from full freeze-out of the ocean, down to a 1 km thick ice shell, and ductile ice viscosity variations from $1 \times 10^{12}$ to $1 \times 10^{21}$ Pa s).

Following a layer description, a solution of the tidal response for a 1D column of material subjected to cyclical forcing is "propagated' from the core to the surface, with a boundary condition that specifies that radial and tangential stresses are zero at the surface (but displacements may be finite) and that the gravitational gradient is contiguous across the surface. This leads to a vector solution for the radial and tangential displacements, radial and tangential stresses, gravitational potential, and a final quantity rarely used in practice known as the gravitational potential stress. These are general solutions that may then be used for any location on the object, and for any type of forcing, such as due to nonzero eccentricity, or non-synchronous spin.

Next, a form of forcing is selected, in this application, the potential equation for the gravitational potential $\Phi_e$ of an object with non-zero eccentricity, as well as both synchronous spin and zero obliquity angle (Wahr et al., 2009).

$$\Phi_e(t, \ r, \ \theta, \phi) = -e \ r^2 \ C \frac{3}{2} \wp_{20}(\theta) \cos nt - e \ r^2 C \frac{1}{4} \wp_{22}(\theta) \ (\cos nt \cos 2\phi + 4 \sin nt \sin 2\phi) \ . \ \text{(A2)}$$

Here $\theta$ is colatitude (0 at the north pole, to $\pi$ at the south pole), $\phi$ is east-longitude (0 at the sub-perturber point), e is eccentricity, r is a radius within the secondary body being evaluated for tides, up to the limit of the surface radius $R_{sec}$ and C is a constant defined as,

$$C = \ \frac{G \ M_P}{a^3} \ . \qquad \text{(A3)}$$

$M_P$ represents the mass of the primary body or tidal pertutrber, Jupiter in this application. The terms $\wp_{20}$ and $\wp_{22}$ represent associated Legendre polynomials. In this case $\Phi_e$ represents the dynamic component of the tidal potential for nonzero-eccentricity, while the static component $\Phi_s$ has been subtracted out. Because the static bulge component causes no temporal changes in the material, it is not reflected in dissipation. The static component may be written (see e.g., Wahr et al., 2009, eqns 2, 3 and associated discussion),

$$\Phi_s(\theta, \phi) = C\frac{1}{2}\wp_{20}(\theta) + C\frac{1}{4}\wp_{22}(\theta)\cos 2\phi \quad . \qquad (A4)$$

Equation A2 is expressed in spherical harmonics as a function of latitude, longitude, and time. Combining this expression with the solution vector as a function of radius, we obtain all component terms of the full stress and strain tensors throughout the object. In our case, we make the standard assumption of an incompressible body, which is quite safe for a low gravity object such as Europa, as it is often made for studies of the Earth itself, with minimal impact on results (see e.g., Sabadini and Vermeersen 2004, Ch 2, Sec 1).

Lastly, the heating in any given parcel of material may be computed at any time as a summation of products of stress and strain tensor terms. Typically, at this step, it is customary to integrate over both spherical volume and time to obtain global orbit averaged solutions (see Eqn 7 of Roberts and Nimmo 2008). In our application, we integrate selectively, in order to obtain results as maps that evolve in time throughout one orbit, or as global totals that evolve in time.

$$E_T(\theta, \phi) = \omega \iint Im(\sigma_{ij})Re(\epsilon_{ij}) - Re(\sigma_{ij})Im(\epsilon_{ij}) \, dVdt \quad . \qquad (A5)$$

where $\epsilon_{ij}$ are the component terms of the strain tensor (9 unique) and $\sigma_{ij}$ are the terms of the stress tensor, $\omega$ is the orbital frequency, dV represents integration by volume, and Re and Im select for the real and imaginary components of the tensor terms following the Fourier domain method of computing viscoelastic tides. The bounds for the second integral in Eqn A5 over time, with corresponding dt, may be selected to encompass any subset of the orbit of interest, such as several even steps in time. Note that this differs significantly from the conventional derivation of the classical tidal equation, as presented for example in Murray & Dermott 1999, Eqns 4.186 to 4.197. The classical derivation begins from the definition of the Quality factor, Q, and therefore sidesteps any explicit introduction of time as a free variable throughout an orbit, precluding any simple ability to observe snapshots of the tidal heating intensity as it varies throughout the orbit by that pathway (The Quality factor derivation invokes only the peak stored energy, which occurs at one unique moment in the orbital cycle). This helps to explain why the coefficient 0.143 discussed in the body text is difficult to determine analytically: as this sub-orbital behavior is resolved numerically through the machinery of: first the propagator matrix formalism; then blending with the tidal potential as in Eqn A2 over latitude, longitude, radius, and time; then lastly blending together 9x2 tensor terms through the integration of Eqn A5. Numerical determination of this sub-orbital behavior is however straightforward by the procedure above, and perhaps easier for others to verify, to find insensitivity of the ~0.143 coefficient to input parameters other than the layer structure.

In this method, a given material rheology is modeled using a complex-valued material compliance (see Table 1 in Henning et al. 2009, or Tables 3-5 in Renaud & Henning 2018). For simplicity in this paper, we use the Maxwell rheology for all layers. While the more advanced Andrade rheology may better model the frequency dependence of the response of ice, rheology variations generally alter only the total magnitude of tidal heating, and not its geometric pattern, because it is applied to laterally homogenous layers. In this sense, given we are not evolving the orbital period in time (and the total dissipation rate for Europa is well-constrained), uncertainty in the selection of rheology is effectively identical to uncertainty in the choice of baseline ice viscosity (which by proxy, is largely controlled by the unknown nature of Europa's ice grain size) (Goldsby & Kohlstedt 2001). Our use of a viscosity of 1e14 Pa s, for a ductile ice layer, if at a temperature at 270 K, would correspond to having a grain size of ~0.35 mm (assuming evaluation at zero pressure, a stress exponent of 1.0, and an activation energy 59400 J/mol for diffusion creep) (following Moore, 2006, Eqn 2 and Table 1).

Supplemental Online Material

Figure 4 of the paper presents seismic activity at 8 points in one orbit of Europa. The seismic activity is a dynamic process that is not captured in the static frames. We animated a seismic catalog for Europa that spans 10-orbital cycles or 35.51 days.